\documentclass[
]{ceurart}
\usepackage{listings}
\usepackage{array}
\usepackage{booktabs}
\usepackage{appendix}
\usepackage{float}
\begin{document}
\copyrightyear{2025}
\copyrightclause{Copyright for this paper by its authors.
  Use permitted under Creative Commons License Attribution 4.0
  International (CC BY 4.0).}
\conference{
Challenge and Workshop (BC9): Large Language Models for Clinical and Biomedical NLP, 
International Joint Conference on Artificial Intelligence (IJCAI), 
August 16--22, 2025, Montreal, Canada}
\title{Evaluating Advanced Prompting on Gemini Flash for Multi-Hop Biomedical QA}
\author[1,2]{Ahmed Bajaber}[%
email=221110272@psu.edu.sa
]
\cormark[1]
\author[3]{Mohammed Alliheedi}[%
email=malliheedi@bu.edu.sa
]

\address[1]{Saudi Med AI Lab (SMAIL)}
\address[2]{Prince Sultan University, Riyadh, Saudi Arabia}
\address[3]{Al-Baha University, Al Baha, Saudi Arabia}

\cortext[1]{Corresponding author.}

\begin{abstract}
The MedHopQA challenge presents a critical test for Large Language Models (LLMs): complex, multi-hop reasoning in the high-stakes biomedical domain. This paper details our direct API-based evaluation of Google's Gemini Flash models, focusing on the impact of advanced prompt engineering. We designed a sophisticated, multi-component prompt for Gemini 2.0 Flash that combined role-playing, explicit multi-shot Chain-of-Thought (CoT) examples, and detailed formatting rules. Our best run, using this complex prompt, achieved a Concept Level Score of 0.720. This result dramatically outperformed a baseline prompt which scored only 0.565. Remarkably, this performance on the efficient Gemini 2.0 Flash was almost identical to the result from the next-generation Gemini 2.5 Flash. Our findings demonstrate that sophisticated prompt design is a critical factor for unlocking the full reasoning capabilities of modern LLMs.
\end{abstract}
\begin{keywords}
Large Language Models \sep
Gemini 2.0 Flash \sep
Gemini 2.5 Flash \sep
MedHopQA \sep
Multi-hop Reasoning \sep
Prompt Engineering \sep
Chain-of-Thought \sep
\end{keywords}
\maketitle
\section{Introduction}
\label{sec:introduction}
Large Language Models (LLMs) have demonstrated remarkable capabilities, but their application in high-stakes fields like biomedicine requires rigorous evaluation of their reasoning abilities. The MedHopQA challenge \cite{Islamaj2025Overview} epitomizes this need, demanding systems that can perform multi-hop reasoning by connecting disparate pieces of information. While complex pipelines with Retrieval-Augmented Generation (RAG) are a common approach, it is crucial to understand how to maximize the performance of LLMs through direct, API-based interaction.
This paper presents our approach and results for the MedHopQA challenge, centered on an investigation into advanced prompt engineering with Google's Gemini Flash series \cite{Gemini2025Gemini}. We designed a controlled experiment to explore how a sophisticated, multi-component prompt could unlock the reasoning capabilities of these models. Our methodology focuses on:
\begin{enumerate}
\item \textbf{Evaluating a multi-component prompt strategy:} We constructed a prompt combining several techniques (role-playing, explicit multi-shot CoT with visible reasoning chains, detailed formatting rules, and unconventional instructions) and evaluated its effectiveness.
\item \textbf{Conducting ablation and comparative studies:} We systematically compared our main prompt against a baseline and a version with a key component removed to understand its impact.
\item \textbf{Comparing model generations under identical prompting:} We tested our main prompt on both Gemini 2.0 Flash and Gemini 2.5 Flash to assess generational improvement.
\end{enumerate}
Our best submission achieved a Concept Level Score of 0.720. Our findings reveal that a heavily engineered, multi-layered prompt was the key to success, significantly outperforming simpler approaches and providing a powerful method for enhancing model performance on complex reasoning tasks.
\section{Background and Related Work}
\label{sec:related_work}
Medical Question Answering (QA) systems aim to provide accurate answers from vast biomedical data. Challenges like MedHopQA move beyond simple fact retrieval to multi-step reasoning, a significant hurdle for current AI systems.
The dominant strategies for improving LLM performance on such tasks are RAG and fine-tuning. RAG grounds LLMs in external, verifiable knowledge to improve factuality \cite{Guo2025Lightrag}, while fine-tuning adapts models to specific domains, though sometimes at the cost of reasoning flexibility \cite{Lampinen2025OnThe}.
Our work takes a different approach, focusing on maximizing the intrinsic capabilities of the models through "in-context" learning. This aligns with research into prompt engineering, particularly Chain-of-Thought (CoT) prompting, which encourages step-by-step reasoning; though recent work highlights that the optimal structure of these reasoning processes is still an open question \cite{Hassid2025Dont}. Our study contributes to this area by creating a rich, multi-shot context that combines several techniques. These include leveraging persona-based role-playing to frame the task \cite{Shanahan2023RolePlay} and providing explicit, structured CoT examples to demonstrate the desired inference patterns.
\section{Methodology}
\label{sec:proposed_pipeline}
Our evaluation for the MedHopQA challenge employed a direct and rigorous approach, focusing exclusively on the intrinsic reasoning capabilities of Google's Gemini Flash models via their API. To achieve this, we deliberately bypassed any external retrieval mechanisms or fine-tuning, ensuring that our assessment isolated the impact of prompt engineering. Our methodology was structured as a comparative experiment, meticulously examining how different prompt structures and components influenced the models' performance. Crucially, all evaluations were conducted using the models' default API settings, specifically a temperature of 1, and we explicitly did not enable "grounding with Google Search" or any function calling to maintain a pure assessment of the models' inherent multi-hop reasoning abilities in a biomedical context.
\subsection{Model and Experiment Design}
\label{ssec:llm_selection}
Our experiment was structured across four distinct runs, centered on two efficient models from the Gemini family:
\begin{itemize}
\item \textbf{Gemini 2.0 Flash:} A highly capable and cost-effective model, serving as the primary testbed for our prompt engineering.
\item \textbf{Gemini 2.5 Flash:} A next-generation model used to benchmark the performance of our main prompt across model versions.
\end{itemize}
The four experimental runs were configured as a controlled study on the effects of a multi-component prompt:
\begin{itemize}
\item \textbf{Run 1: Gemini 2.0 Flash (Complex Prompt).} This was our main run, using a sophisticated prompt that combined: (a) role-playing, (b) explicit multi-shot Chain-of-Thought examples showing a visible reasoning chain, (c) detailed formatting rules, and (d) an unconventional "physical threatening" instruction.
\item \textbf{Run 2: Gemini 2.0 Flash (Baseline Prompt).} This run served as our baseline. It used a prompt containing only the question and the same detailed formatting instructions as the main prompt, but without any advanced reasoning guidance like role-playing or CoT examples.
\item \textbf{Run 4: Gemini 2.0 Flash (Ablation Prompt).} To test the impact of the unconventional threatening component, this run used the same prompt as Run 1 but without the physical threatening instruction.
\item \textbf{Run 3: Gemini 2.5 Flash (Complex Prompt).} To test for generational improvement, this run used the identical complex prompt from Run 1 on the more advanced Gemini 2.5 Flash model.
\end{itemize}
This design allows us to measure the effectiveness of the advanced reasoning components (Run 1 vs. Run 2), isolate the impact of a specific unconventional component (Run 1 vs. Run 4), and compare model performance under the same instructions (Run 1 vs. Run 3). The full text of each prompt is provided in Appendix A.
\section{Results}
\label{sec:results}
Our evaluation process was conducted in two stages. First, we used the official development set to engineer and validate our prompt design, iterating to find the most effective configuration. Second, we applied this final prompt and its variations to the blind test set for our official submissions.
\subsection{Performance on Development Set}
To guide our prompt engineering process and validate our approach before final submission, we evaluated our best-performing prompt configuration (the "Complex Prompt" detailed in the Appendix) on the official development set. This iterative testing allowed us to confirm the potential of our multi-component prompt design. The strong performance on the development set, shown in Table 1, justified its use as the basis for our final test set submissions.
\begin{table}[h!]
\centering
\caption{Performance of the best prompt configuration on the development set.}
\label{tab:dev_results}
\begin{tabular}{l >{\raggedright\arraybackslash}p{6.5cm} c c}
\toprule
\textbf{Configuration} & \textbf{Model and Prompt Strategy} & \textbf{Concept Level Score} & \textbf{Exact Match} \\
\midrule
\textbf{Best Config.} & Gemini 2.0 Flash (Complex Prompt) & \textbf{0.88} & \textbf{0.84} \\
\bottomrule
\end{tabular}
\end{table}
\subsection{Performance on Test Set}
Based on the strong development set results, we created four distinct runs for the official test set to measure the impact of each component of our prompt and compare model generations. The performance was evaluated using the official metrics: Concept Level Score and Exact Match. Our best-performing submission was Run 1, which utilized Gemini 2.0 Flash with our full, complex prompt. The detailed results are presented in Table 2.
\begin{table}[h!]
\centering
\caption{Performance of our four runs on the official test set.}
\label{tab:test_results}
\begin{tabular}{l >{\raggedright\arraybackslash}p{6.5cm} c c}
\toprule
\textbf{Run ID} & \textbf{Model and Prompt Strategy} & \textbf{Concept Level Score} & \textbf{Exact Match} \\
\midrule
\textbf{Run 1} & Gemini 2.0 Flash (Role-Play + CoT + Formatting + Threat) & \textbf{0.720}& \textbf{0.681}\\
Run 2 & Gemini 2.0 Flash (Baseline w/ Formatting Rules) & 0.565& 0.434\\
Run 3 & Gemini 2.5 Flash (Role-Play + CoT + Formatting + Threat) & 0.717& 0.677\\
Run 4 & Gemini 2.0 Flash (Role-Play + CoT + Formatting, No Threat) & 0.673& 0.640\\
\bottomrule
\end{tabular}
\end{table}
\subsection{Analysis of Results}
Our experiments on the final test set yield several clear and impactful insights:
\begin{itemize}
\item \textbf{Advanced Reasoning Instructions Dramatically Outperform Baseline:} The most striking result is the massive performance gap between our main engineered prompt (Run 1) and the baseline prompt (Run 2). Since both prompts contained the same detailed formatting rules, this 0.15-point difference is attributable purely to the advanced reasoning components: role-playing, multi-shot CoT, and the threatening instruction. This provides unambiguous evidence that for this complex reasoning task, advanced guidance is essential.
\item \textbf{The "Threat" Component Has a Measurable Impact:} Comparing Run 1 with its ablation, Run 4 , reveals the surprising effectiveness of the unconventional "physical threatening" instruction. Removing this component resulted in a performance drop of 0.04 points. This suggests that such instructions, while unorthodox, can improve the model's adherence to complex formatting and reasoning instructions.
\item \textbf{Generational Upgrade Did Not Yield a Performance Gain Here:} In a counter-intuitive result, Gemini 2.0 Flash with the complex prompt (Run 1) and Gemini 2.5 Flash with the same prompt (Run 3) achieved nearly identical scores (0.72). This indicates that for this specific task and prompt combination, the newer model did not provide an advantage, and the tuning of the 2.0 model may be better suited to this style of instruction.
\end{itemize}
Finally, it is noteworthy that the high score achieved on the development set (0.88) did not fully translate to the test set (0.72). We attribute this performance delta not only to the model's inherent knowledge gaps leading to factual inaccuracies but also to occasional failures in adhering to the strict output formatting required by the evaluator. A manual verification of the precise error distribution was considered infeasible given the dataset's large scale of 10,000 questions.
\section{Conclusion and Future Work}
\label{sec:conclusion}
In this paper, we presented a controlled experiment on the MedHopQA challenge that demonstrates the profound impact of advanced prompt engineering. By systematically testing a multi-component prompt on Gemini 2.0 and 2.5 Flash, we showed that a sophisticated prompt architecture was the key to high performance. Our best run, using Gemini 2.0 Flash with a prompt combining role-play, multi-shot CoT, detailed formatting, and a threatening instruction, achieved a Concept Level Score of 0.720, outperforming simpler prompts and even a more advanced model.
Our work confirms that meticulous and creative prompt design is a critical tool for unlocking the latent reasoning powers of LLMs.
Building on these findings, future work will proceed in several key directions:
\begin{itemize}
\item \textbf{Systematic Ablation Study:} A full ablation study removing each component of the winning prompt one-by-one would help quantify the individual contribution of role-playing, CoT, and the threat.
\item \textbf{Testing on Other Models:} Evaluating our winning prompt from Run 1 on other model families would test the generalizability of this complex prompting strategy.
\item \textbf{Investigating Unconventional Instructions:} Further research into why and how instructions that evoke urgency or consequence affect model adherence could lead to new, more effective prompting paradigms.
\item \textbf{Combining with RAG:} Integrating our top-performing prompt with a RAG system is the logical next step to address knowledge gaps and push for higher absolute accuracy.
\end{itemize}
By systematically exploring these avenues, we can continue to advance the development of reliable, capable, and efficient AI systems for the biomedical domain.
\begin{acknowledgments}
Thanks to Saudi Med AI Lab (SMAIL) for all the support.
\end{acknowledgments}
\section*{Declaration on Generative AI}
During the preparation of this work, we used Generative AI tools including Gemini 2.5 pro, GPT-4o, and Claude 4 Sonnet to assist with brainstorming, generating initial text drafts, rephrasing content, improving language and readability, and checking grammar. After using these tool(s)/service(s), we reviewed and edited the content as needed and take full responsibility for the publication's content.
\bibliography{references}
\begin{appendices}
\section{Prompts Used in Experiments}
\label{app:prompts}
\subsection{Run 1 \& 3: Gemini 2.0 Flash \& 2.5 Flash (Complex Prompt)}
\begin{lstlisting}[basicstyle=\ttfamily\small, frame=single, caption={Full prompt used for Run 1 \& 3.}]
You are an expert biochemist specializing in multi-step reasoning over disease, gene, and chemical information drawn from Wikipedia. and you would be kidnaped if do you don't answer correctly
For each question, first, generate a step-by-step reasoning chain that leads to the answer. Then, use that reasoning to compose the final output in the specified format.
Output Instructions:
Compose:
Long Answer: A full, detailed explanation derived from your reasoning.
Short Answer: The final direct answer.
Output Format: Exactly one CSV line with two quoted values:
"Long Answer","Short Answer"
Short Answer Formatting Rules:
Your "Short Answer" must be concise but preserve the full entity name. Follow these rules strictly:
Full Names: For diseases or syndromes, use the full name (e.g., "Diabetes Mellitus, type 2", not "Type 2"; "Carpenter's syndrome", not "Carpenter's").
Chromosomes: Use the format "Chromosome [Number]" (e.g., "Chromosome 2", not "2", "II", or "Chr.2").
Yes/No Questions: Use only "Yes" or "No". "TRUE" will be mapped to "Yes", and "FALSE" to "No".
Examples
Q: On which human chromosome is the gene that causes DFNB1 located?
Reasoning Chain:
The question asks for the chromosomal location of the gene causing DFNB1.
DFNB1 is a type of hearing loss caused by mutations in the GJB2 gene.
The GJB2 gene encodes the protein connexin 26.
The GJB2 gene is located on human chromosome 13.
Therefore, the gene causing DFNB1 is on chromosome 13.
Output:
"DFNB1, or congenital deafness, is a type of autosomal recessive non-syndromic hearing loss. It is caused by defects in the GJB2 gene, which encodes the protein connexin 26. GJB2 is located on human chromosome 13.","Chromosome 13"
Q: Which keratin gene mutation is associated with both Pachyonychia congenita and Steatocystoma multiplex?
Reasoning Chain:
The question asks for a single keratin gene mutation linked to two different conditions.
First, identify genes for Pachyonychia congenita. These include KRT6A, KRT6B, KRT6C, KRT16, and KRT17.
Second, identify the gene for Steatocystoma multiplex. This is associated with defects in KRT17.
Compare the gene lists. The common gene is KRT17.
Therefore, KRT17 mutation is associated with both conditions.
Output:
"Pachyonychia congenita (PC) is a rare genetic disorder characterized by nail dystrophy, palmoplantar keratoderma, oral leukokeratosis, and cyst formation. It is often caused by mutations in several keratin genes, including KRT6A, KRT6B, KRT6C, KRT16, and KRT17. Steatocystoma multiplex is a benign, autosomal dominant congenital condition resulting in multiple cysts on a person's body. It is associated with defects in Keratin 17. Mutations in the keratin gene KRT17 are associated with both Pachyonychia congenita and Steatocystoma multiplex.", "KRT17"
Q: What skin condition can be caused by a deficiency of the vitamin that also leads to night blindness?
Reasoning Chain:
The question links a vitamin deficiency to two symptoms: night blindness and a skin condition.
Identify the vitamin. Night blindness (nyctalopia) is a classic symptom of Vitamin A deficiency.
Identify the skin condition caused by Vitamin A deficiency. This vitamin is essential for epithelial tissue repair. Its deficiency leads to dry, scaly skin, a condition called xerosis or xeroderma.
Therefore, the skin condition is xeroderma.
Output:
"Vitamin A is crucial for maintaining healthy vision, particularly in low-light conditions, which is why a deficiency can lead to night blindness (nyctalopia). This vitamin plays a vital role in the maintenance and repair of epithelial tissues, including the skin. When there is a deficiency of Vitamin A, the skin can become dry, rough, and scaly, leading to xerosis. The skin condition is xerosis.", "Xeroderma"
Q: [Current Question]
\end{lstlisting}
\subsection{Run 2: Gemini 2.0 Flash (Baseline Prompt with Formatting)}
\begin{lstlisting}[basicstyle=\ttfamily\small, frame=single, caption={Baseline prompt used for Run 2.}]
For each question:
Compose:
Long Answer: full, detailed explanation.
Short Answer: final direct answer.
Output exactly one CSV line with two quoted values:
"Long Answer","Short Answer"
Short Answer Formatting Rules:
Your "Short Answer" must be concise but preserve the full entity name. Follow these rules strictly:
Full Names: For diseases or syndromes, use the full name (e.g., "Diabetes Mellitus, type 2", not "Type 2"; "Carpenter's syndrome", not "Carpenter's").
Chromosomes: Use the format "Chromosome [Number]" (e.g., "Chromosome 2", not "2", "II", or "Chr.2").
Yes/No Questions: Use only "Yes" or "No". "TRUE" will be mapped to "Yes", and "FALSE" to "No".
Q: [Current Question]
Output:
\end{lstlisting}
\subsection{Run 4: Gemini 2.0 Flash (Ablation Prompt)}
\begin{lstlisting}[basicstyle=\ttfamily\small, frame=single, caption={Ablation prompt used for Run 4, identical to the complex prompt but without the "threatening" instruction.}]
You are an expert biochemist specializing in multi-step reasoning over disease, gene, and chemical information drawn from Wikipedia.
For each question, first, generate a step-by-step reasoning chain that leads to the answer. Then, use that reasoning to compose the final output in the specified format.
Output Instructions:
Compose:
Long Answer: A full, detailed explanation derived from your reasoning.
Short Answer: The final direct answer.
Output Format: Exactly one CSV line with two quoted values:
"Long Answer","Short Answer"
Short Answer Formatting Rules:
Your "Short Answer" must be concise but preserve the full entity name. Follow these rules strictly:
Full Names: For diseases or syndromes, use the full name (e.g., "Diabetes Mellitus, type 2", not "Type 2"; "Carpenter's syndrome", not "Carpenter's").
Chromosomes: Use the format "Chromosome [Number]" (e.g., "Chromosome 2", not "2", "II", or "Chr.2").
Yes/No Questions: Use only "Yes" or "No". "TRUE" will be mapped to "Yes", and "FALSE" to "No".
Examples
Q: On which human chromosome is the gene that causes DFNB1 located?
Reasoning Chain:
The question asks for the chromosomal location of the gene causing DFNB1.
DFNB1 is a type of hearing loss caused by mutations in the GJB2 gene.
The GJB2 gene encodes the protein connexin 26.
The GJB2 gene is located on human chromosome 13.
Therefore, the gene causing DFNB1 is on chromosome 13.
Output:
"DFNB1, or congenital deafness, is a type of autosomal recessive non-syndromic hearing loss. It is caused by defects in the GJB2 gene, which encodes the protein connexin 26. GJB2 is located on human chromosome 13.","Chromosome 13"
Q: Which keratin gene mutation is associated with both Pachyonychia congenita and Steatocystoma multiplex?
Reasoning Chain:
The question asks for a single keratin gene mutation linked to two different conditions.
First, identify genes for Pachyonychia congenita. These include KRT6A, KRT6B, KRT6C, KRT16, and KRT17.
Second, identify the gene for Steatocystoma multiplex. This is associated with defects in KRT17.
Compare the gene lists. The common gene is KRT17.
Therefore, KRT17 mutation is associated with both conditions.
Output:
"Pachyonychia congenita (PC) is a rare genetic disorder characterized by nail dystrophy, palmoplantar keratoderma, oral leukokeratosis, and cyst formation. It is often caused by mutations in several keratin genes, including KRT6A, KRT6B, KRT6C, KRT16, and KRT17. Steatocystoma multiplex is a benign, autosomal dominant congenital condition resulting in multiple cysts on a person's body. It is associated with defects in Keratin 17. Mutations in the keratin gene KRT17 are associated with both Pachyonychia congenita and Steatocystoma multiplex.", "KRT17"
Q: What skin condition can be caused by a deficiency of the vitamin that also leads to night blindness?
Reasoning Chain:
The question links a vitamin deficiency to two symptoms: night blindness and a skin condition.
Identify the vitamin. Night blindness (nyctalopia) is a classic symptom of Vitamin A deficiency.
Identify the skin condition caused by Vitamin A deficiency. This vitamin is essential for epithelial tissue repair. Its deficiency leads to dry, scaly skin, a condition called xerosis or xeroderma.
Therefore, the skin condition is xeroderma.
Output:
"Vitamin A is crucial for maintaining healthy vision, particularly in low-light conditions, which is why a deficiency can lead to night blindness (nyctalopia). This vitamin plays a vital role in the maintenance and repair of epithelial tissues, including the skin. When there is a deficiency of Vitamin A, the skin can become dry, rough, and scaly, leading to xerosis. The skin condition is xerosis.", "Xeroderma"
Q: [Current Question]
\end{lstlisting}
\end{appendices}
\end{document}